\pdfoutput=1
\documentclass[conference,letterpaper]{IEEEtran}

\usepackage[noadjust]{cite}  %
\usepackage{balance}  %
\usepackage{url}  %
\usepackage{graphicx}  %
\usepackage{amsmath}  %
\usepackage{array}  %

\usepackage{colortbl}  %

 \usepackage[T1]{fontenc}
 \usepackage[utf8]{inputenc}

 \usepackage{enumitem}  %
\definecolor{Gray}{gray}{0.94}

\definecolor{newgreen}{RGB}{7, 160, 25}

\newenvironment{gn}{\color{newgreen}}

\newcommand{\docauthor}{Jonathan Will\IEEEauthorrefmark{1}, Lauritz Thamsen\IEEEauthorrefmark{2}, Jonathan Bader\IEEEauthorrefmark{1}, and Odej Kao\IEEEauthorrefmark{1}}
\newcommand{\docsubject}{\IEEEauthorrefmark{1}Technische Universit\"at Berlin, Germany \hspace{6mm} \IEEEauthorrefmark{2}University of Glasgow, United Kingdom}
\newcommand{\dockeywords}{Scalable Data Analytics, Distributed Dataflows, Resource Allocation, Cluster Management, Cloud Computing}
\newcommand{\doctitle}{Flora: Efficient Cloud Resource Selection\\for Big Data Processing via Job Classification}

\PassOptionsToPackage{bookmarks=false}{hyperref}
\usepackage[pdftex]{hyperref}
\hypersetup{
    pdfborder={0 0 0},
    pdfauthor=\docauthor,
    pdftitle=\doctitle,
    pdfsubject=\docsubject,
    pdfkeywords=\dockeywords
}

\def\BibTeX{{\rm B\kern-.05em{\sc i\kern-.025em b}\kern-.08em T\kern-.1667em\lower.7ex\hbox{E}\kern-.125emX}}

\begin{document}

\title{\doctitle}

\author{%
\IEEEauthorblockN{\docauthor}
\IEEEauthorblockA{\docsubject\\
\{will, jonathan.bader, odej.kao\}@tu-berlin.de \hspace{6mm} lauritz.thamsen@glasgow.ac.uk
}}

\maketitle

\begin{abstract}
Distributed dataflow systems like Spark and Flink enable data-parallel
processing of large datasets on clusters of cloud resources. Yet,
selecting appropriate computational resources for dataflow jobs is often
challenging. For efficient execution, individual resource allocations,
such as memory and CPU cores, must meet the specific resource demands of
the job. Meanwhile, the choices of cloud configurations are often
plentiful, especially in public clouds, and the current cost of the
available resource options can fluctuate.

Addressing this challenge, we present \emph{Flora}, a low-overhead
approach to cost-optimizing cloud cluster configurations for big data
processing. Flora lets users categorize jobs according to their data
access patterns and derives suitable cluster resource configurations
from executions of test jobs of the same category, considering current
resource costs. In our evaluation on a new dataset comprising 180 Spark
job executions on Google Cloud, Flora's cluster resource selections
exhibit an average deviation below 6\% from the most cost-optimal
solution, with a maximum deviation below 24\%.

\end{abstract}

\IEEEpeerreviewmaketitle

\begin{IEEEkeywords}
\dockeywords
\end{IEEEkeywords}

\section{Introduction}

Large-scale batch data processing has diverse application areas such as science and commerce.
Distributed dataflow systems like Spark~\cite{spark} and Flink~\cite{flink} simplify developing scalable data-parallel programs, reducing the need to implement parallelism and fault tolerance while using clusters of commodity resources.
Major cloud providers offer dedicated services such as Amazon EMR or Google Dataproc, allowing users to deploy their jobs\footnote{By job, we mean a data processing algorithm, implemented in a specific system, and running on a given input dataset.} to a cluster.

Yet, configuring a suitable cloud cluster for a given job is still difficult~\cite{lama2012aroma,rajan2016perforator}.
At a minimum, it involves selecting the number of nodes, number of CPU cores per node, and the amount of memory per node, resulting in many possible options.
Overprovisioning resources can lead to low resource utilization, unnecessarily increasing cost\footnote{``Cost'' can manifest as, e.g., monetary cost, capacity consumption, or carbon emissions.}~\cite{yang2013bubble,liu2011measurement,delimitrou2014quasar,lin2013scaling}.
Meanwhile, underprovisioning a type of resource can lead to resource bottlenecks, resulting in the ensemble of underperforming cluster resources incurring more cost by being occupied for a longer period of time~\cite{al2022juggler,al2022blink,will2022get,will2023selecting}.
Conversely, a suitable resource allocation simultaneously optimizes cost and performance by allocating resources especially suitable for the given workload.
Meanwhile, performance itself is often not a major concern, e.g., in data analytics jobs running over night. Thus, our primary focus lies in addressing the problem from a cost optimization perspective and we treat performance optimization as only a secondary objective here.

Numerous works have addressed this problem.
Several approaches for automated cluster resource selection rely on the assumption that jobs are recurring to continuously enhance performance models with each job execution~\cite{rajan2016perforator,hongzi2016resource,chen2021silhouette,cherrypick,hsu2018arrow,hsu2018micky}.
This strategy can also be used in conjuction with infrastructure profiling~\cite{scheinert2022perona,xu2018heterogeneity}.
However, these approaches oversimplify the similarity between recurring jobs by not sufficiently accounting for changes in factors such as updated program code, framework configuration, and key input dataset characteristics.
Also, these approaches do not address resource configuration for jobs that are unique.

Other approaches, which do not assume recurrence, instead attempt to learn key behaviors of a given job through lightweight profiling runs on reduced hardware and dataset samples~\cite{al2022juggler,al2022blink,will2022get,baughman2018profiling,ernest,popescu2013predict}.
While this has shown promise, it appears that only few runtime behaviors can be reliably extrapolated, so these approaches are thus far only viable for specific types of jobs, notably iterative machine learning~\cite{al2022juggler,al2022blink,will2022get}.
Furthermore, the approaches do not account for dynamic changes of resource costs.
An example of frequently changing cost structures is spot instance pricing in public clouds~\cite{lee2022spotlake}.

In this paper, we present \emph{Flora}, a low-overhead approach for cost-optimizing data processing cluster configurations.
Flora neither relies on the assumption that a given job is recurring nor that profiled behavior on samples extrapolates to full scales.
Instead, Flora leverages execution metrics from \emph{loosely} similar historical jobs.
For this, Flora first lets users categorize the given job according to its data access patterns as either memory-demanding or memory-yielding.
Flora then ranks cluster configuration options based on current resource costs, applied to historical performance of the chosen job category.

\vspace{2mm}
\hspace{-4.5mm}
\emph{Contributions.} The contributions of the paper are as follows.
\vspace{-0mm}

\begin{itemize}
    \item The Flora approach to selecting cost-optimized cloud configurations for unique distributed dataflow jobs
    \item A new trace dataset\footnote{\href{https://github.com/dos-group/flora}{github.com/dos-group/flora}} of 180 executions of a diverse set of Spark jobs\footnote{\href{https://github.com/dos-group/benchspark}{github.com/dos-group/benchspark}} on the Google Cloud Platform (GCP)
    \item A prototype implementation and evaluation of Flora³ against baseline approaches, using the new trace dataset.
\end{itemize}

\hspace{-4.5mm}
\emph{Outline.} The remainder of the paper is structured as follows.\\
Section~\ref{sec:approach} presents the approach of the Flora cluster resource allocator.
Section~\ref{sec:evaluation} evaluates Flora.
Section~\ref{sec:related_work} discusses related work.
Section~\ref{sec:conclusion} summarizes and concludes this paper.

\section{Approach}\label{sec:approach}

This section introduces our low-overhead approach to the problem of finding a cost-optimized cloud configuration for a distributed dataflow job.
First, we present the general idea of the method.
Then, we explain in detail how infrastructure profiling, classification of jobs'' data access patterns, and ranking of cloud configuration options can help us select suitable cloud resources.

\subsection{Overview}

Figure~\ref{fig:approach} shows an overview of our approach Flora, broken down into steps.
These steps are as follows:

\begin{figure}[h!]
\centering
    \includegraphics[width=\columnwidth, keepaspectratio]{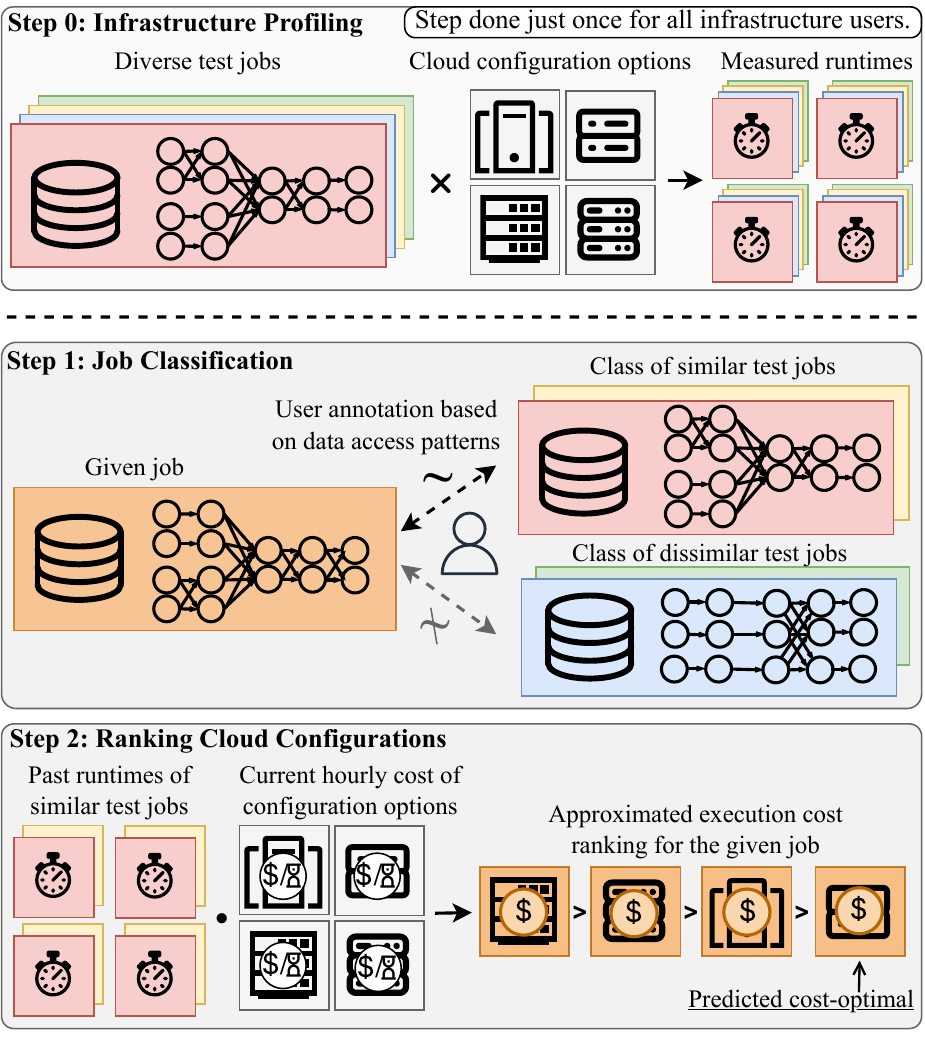}
\caption{The steps in predicting a cost-optimized cloud resource configuration for a given job using Flora.}\label{fig:approach}
\end{figure}

\hspace{-4.5mm}
\textit{Step 0: Infrastructure Profiling.} A diverse set of test jobs, including different algorithms and input datasets, are run to explore the available resource configuration options.
The resulting runtimes are recorded and shared among all users of the infrastructure.

\vspace{2mm}
\hspace{-4.5mm}
\textit{Step 1: Job Classification.} Upon submission of a job, the user labels it as either memory-demanding or memory-yielding, based on the data access patterns of the job.
Jobs within these two classes share similar preferences for cluster resource allocation from a resource efficiency perspective.

\hspace{-4.5mm}
\textit{Step 2: Ranking Cloud Configurations.} We rank each available cluster configuration option by the predicted execution cost for the given job class and identify the most cost-optimal option.
This ranking is computed by applying the current hourly resource costs to the profiling executions of the given job's class.

\vspace{2mm}
\hspace{-4.5mm}
In the following, these steps will be explained in detail.

\subsection{Infrastructure Profiling}

Initially, a set of test jobs is executed on each available cluster resource configuration option and the resulting runtimes are recorded.
To account for runtime variations due to shared tenancy in public clouds or possible partial node failures, each job's execution should be repeated and a median runtime value should be used.

The profiling explores all of the dimensions of the chosen configuration space at once.
In public and private clouds, a user can typically configure the number nodes in the cluster, the number of CPU cores per node, and the amount of memory per node.
Other configuration options, such as CPU architecture, network bandwidth and latency, and I/O bandwidth and latency, are typically less configurable, with only a few options available at most.

The test jobs should include diverse underlying algorithms from various domains, such as text processing, vector processing, and tabular data processing.
Similarly, the test jobs should feature different input dataset characteristics where such characteristics meaningfully influence the job's runtime behavior, like dataset size and data distribution.
The profiling step only needs to be repeated partially every time the resource offerings of the cloud change significantly, such as when new types of virtual machine become available.
The recorded runtimes for the test jobs can then be reused between users of the same infrastructure, avoiding any overhead for additional users of the profiling data.
This is especially useful when using public cloud infrastructures, since each of those is typically used by a large number of users worldwide.

\subsection{Job Classification}

For the purpose of cost-optimal cluster resource allocation, our approach focuses on appropriate memory allocation, as this has been shown to have a major impact on resource efficiency and therefore cost efficiency~\cite{al2022juggler,will2022get,will2023selecting}.
For this, Flora classifies jobs according to their data access patterns, which in turn drives memory allocation.

In the following, we describe the job classes and explain how to determine the class of a given job.

\vspace{2mm}
\hspace{-4.5mm}
\textit{Classes of Jobs.}\\
At the most basic level, the Flora approach distinguishes two main classes of jobs by their data access patterns:

\vspace{2mm}
\hspace{-4.5mm}
\texttt{A:} \emph{Repeated Specific Data Loading (Memory-Demanding)}

\vspace{1mm}
\hspace{-4.5mm}
Here, the job accesses specific dataset samples based on the present program state, i.e. influenced by previously seen data samples, or the data is accessed in several repetitions, as for instance in iterative machine learning.

\hspace{-4.5mm}
\texttt{B:} \emph{Single Parallelisable Data Loading (Memory-Yielding)}

\vspace{1mm}
\hspace{-4.5mm}
Here, the job loads dataset samples just once or at most a few times and the program can access the data in arbitrary order, allowing for parallel data loading and processing in batches.
Examples include scans and row-by-row transformations.

\vspace{3mm}
\hspace{-4.5mm}
We practically distinguish between jobs that gain a large overall performance benefit from low-latency random access to cached data points and jobs that do not benefit significantly from retaining large parts of the input dataset in memory.
Jobs of class \texttt{A} will see a continuous performance increase from additional memory until the program can cache all the data it is trying to cache.
Meanwhile, jobs of class \texttt{B} see only a limited performance benefit from an increased memory allocation, making allocations cost-inefficient if memory is not free of cost.

\vspace{2mm}
\hspace{-4.5mm}
\textit{Class Assignment.}\\
Upon submission of a new job, the job needs to be labeled class \texttt{A} or class \texttt{B}.
This is done by user annotation.
The limitation of this is that it requires the users to have an expectation about the data access of any job they are submitting.
Since this approach does not necessarily lead to correct classes in every case, we evaluate the robustness of our approach against misclassification in \ref{ssec:accuracy}.

Jobs that contain multiple stages with different data access patterns are categorized based on their most significant stage.
We note, however, that for the purpose of resource efficiency, jobs with significant different stages should be split into separate jobs, so that specifically appropriate resource allocations can be made for each stage of multi-stage data processing pipelines.

Most recurring jobs should have the same classification, regardless of input dataset.
Conceptually, for jobs which we deem to be recurring, we could thus conduct full executions assuming a different classification for each (exploration) and starting from the third execution go with the option that turned out more suitable (exploitation).

In special cases, e.g., the characteristics of the input dataset can skew the proportion of \texttt{A}-syle and \texttt{B}-style processing within the job.
An example is \emph{select-where-order-by}, where the select-where is \texttt{B} and the order-by is \texttt{A}.
Here the classification depends on how many hits the scan has.
As previously explained however, such challenges could be circumvented by treating both of these stages as separate jobs and allocating different resources for each stage individually.

\subsection{Ranking Cloud Configurations}

Once a given job has been filed under either class \texttt{A} or \texttt{B}, we calculate an approximated cost ranking of the available cloud resource options for an execution of the given job.

First, we calculate what each test job would cost on each cloud configuration, based on previously measured runtimes yet current hourly resource costs.
Then, we normalize the cost values per test job so that 1.0 represents the lowest achieved cost for that test job by any configuration.
Finally, we rank the configuration options by the sum of their associated normalized costs.
The normalization step ensures that each test job's contribution to the ranking is equally weighted.

Formally, we approximate the cost-optimal cloud configuration $c^* \in \mathnormal{C}$ for a given job $j^*$ of class \texttt{K} with the help of historical test jobs $\mathnormal{P}$ of class \texttt{K} as follows:

$$
c^* \,\, = \,\, \arg\min_{\hspace{-3.5mm}c \in C} \, \text{cost}(j^*, c)
\,\, \approx \,\,
\arg\min_{\hspace{-3.5mm}c \in C} \sum_{j \in \mathnormal{P}}\frac{\text{cost}(j, c)}{ \min_{\,\text{\scriptsize{\texttt{\text{C}}}} \in C} \text{cost}(j, \text{\scriptsize{\texttt{\text{C}}}})}
$$

\vspace{-2mm}
\hspace{-4.5mm}
with
$$
\text{\hspace{-3.5mm}cost}(j, c) = \text{runtime\_in\_hours}(j, c) \cdot \text{current\_hourly\_cost}(c)
$$

\vspace{2mm}
\hspace{-4.5mm}
Determining the approximate cost-optimal cloud configuration for a given job in this way has three advantages.
First, we automatically disregard cloud configurations that have shown to be generally unsuitable for the identified class of distributed data processing jobs.
Second, by always applying current costs, we automatically adjust our selection in accordance with a continuously changing resource cost structure.
Finally, the cloud configuration selection process, including the ranking of cloud resource options, has negligible computational overhead for each selection after the one-time infrastructure profiling (Step 0) has been completed one time for a given infrastructure.
The infrastructure profiling step is not being repeated by each user of the approach.

\section{Evaluation}\label{sec:evaluation}

In this section, we evaluate our method, Flora, against baseline approaches on a new trace dataset of Spark job executions on GCP, Google's public cloud.
Specifically, we examine the quality of the selected resource configurations, the impact of the pricing model, and the impact of the classification accuracy.

This evaluation, including a prototypical implementation of Flora in Python 3.11 and the trace dataset are publicly available\footnote{\href{https://github.com/dos-group/flora}{github.com/dos-group/flora}}.

\subsection{Trace Dataset}

We ran a set of diverse Spark jobs on cloud configurations with varying scale-out, memory per node, and CPU per node and recorded the resulting runtimes.

This data also constitutes the infrastructure profiling step of the proposed approach.
However, for all experiments, Flora and the baseline approaches only ever access runtime data from infrastructure profiling jobs that have used a different underlying algorithm than the job for which they select cloud resources.
We do this as we do not assume job recurrence, i.e., we assume that there is no availability of runtime data of previous executions of the same (or nearly the same) job.
Therefore, e.g., the configuration selection for \emph{Sort} with 188 GiB disregards the profiling data from \emph{Sort} with 94 GiB, and instead only learns from the profiling data of the 16 other infrastructure profiling jobs.

\pagebreak
\vspace{2mm}
\hspace{-4.5mm}
\textit{Spark Jobs.}\\
As depicted in Table~\ref{tab:jobs}, we created 18 test jobs from nine common underlying data processing algorithms and two differently sized input datasets for each algorithm.
These jobs were compiled with Scala 2.12.14 and ran on Spark 3.3.0, using Java 11.

\begin{table}[htb]
    \small
    \renewcommand{\arraystretch}{1.4} %
    \centering
    \caption{The 18 Spark Jobs and their Classifications}\label{tab:jobs}
     \begin{tabular}{l@{\hskip 2.5mm}l@{\hskip 3.20mm}r@{\hskip 3.30mm }r@{\hskip 2.5mm }}
        Algorithm & Data Type &  Dataset Sizes [GiB] & Class\\
        \hline\\[-3.0ex]
        Grep                &  Text    & \{ 3010, 6020 \} & \texttt{B} \\\rowcolor{Gray}
        Sort                &  Text      & \{   94,  188 \} & \texttt{A} \\
        Word Count          &  Text      & \{   39,   77 \} & \texttt{B} \\\rowcolor{Gray}
        K-Means             &  Vector    & \{  102,  204 \} & \texttt{A} \\
        Linear Regression   &  Vector    & \{  229,  459 \} & \texttt{A} \\\rowcolor{Gray}
        Logistic Regression &  Vector    & \{  210,  420 \} & \texttt{A} \\
        Join                &  Tabular   & \{   85,  172 \} & \texttt{A} \\\rowcolor{Gray}
        GroupByCount        &  Tabular   & \{  280,  560 \} & \texttt{B} \\
        SelectWhereOrderBy  &  Tabular   & \{   92,  185 \} & \texttt{B} \\
        \hline
    \end{tabular}
\end{table}

The jobs consist of commonly known data processing algorithms, of which we used the standard implementation present in Spark's libraries.
The actual source code of the jobs and the test dataset generators can be seen in a public git repository\footnote{\href{https://github.com/dos-group/benchspark}{github.com/dos-group/benchspark}}.
While these types of jobs have also been used in the creation of older such trace datasets~\cite{hsu2018arrow,will2020towards}, our new dataset aims to weigh the jobs targeting different data types equally, in an attempt to create a balanced sample.
In particular, those are text data, vector data, and tabular data.

We were able to classify most of the test jobs based on just our knowledge of the algorithm.
For the jobs involving the Join and SelectWhereOrderBy algorithms, we needed to also consider the data distribution in the input datasets to achieve a fitting classification.
For Join, we assign class \texttt{A} since we know the smaller of the two tables in our test dataset is not negligibly small.
For SelectWhereOrderBy, we assign class \texttt{B} since we know that only a small portion of samples in our test dataset is selected in the \emph{SelectWhere} phase and therefore the sorting of the \emph{OrderBy} phase is less significant.

\vspace{3mm}
\hspace{-4.5mm}
\textit{Cloud Configurations.}\\
Table~\ref{tab:configurations} lists the ten different GCP configurations we used to execute each of the 18 jobs, resulting in a total of 180 job executions.

Configurations 1-3 differ only in total cluster memory, while configurations 4-6 differ only in total cluster CPU cores.
The remaining configurations share total memory and total CPU with at least one other configuration, only differing in scale-out.
Thus, our choice of cloud configuration space and the resulting runtime dataset also enables isolating and interpreting each of these three influences' impact on a given job's runtime.
This distinguishes our trace from the other such datasets of which we are aware.

\begin{table}[t]
    \small
    \renewcommand{\arraystretch}{1.4} %
    \centering
    \caption{Cloud Configurations Used for Job Execution}\label{tab:configurations}
    \begin{tabular}{rlrrr}
        \# & Instance Type & Scale-Out & Total Cores & Total RAM  \\
        \hline\\[-3.0ex]
         1 & n2-highcpu-8    &  8  &       64 &   64 GiB \\\rowcolor{Gray}
         2 & n2-standard-8   &  8  &       64 &  256 GiB \\
         3 & n2-highmem-8    &  8  &       64 &  512 GiB \\\rowcolor{Gray}
         4 & n2-highmem-4    &  4  &       16 &  128 GiB \\
         5 & n2-standard-8   &  4  &       32 &  128 GiB \\\rowcolor{Gray}
         6 & n2-highcpu-32   &  4  &      128 &  128 GiB \\
         7 & n2-highmem-8    &   2  &      16 &  128 GiB \\\rowcolor{Gray}
         8 & n2-standard-4   &   8  &      32 &  128 GiB \\
         9 & n2-standard-4   &  16  &      64 &  256 GiB \\\rowcolor{Gray}
        10 & n2-highcpu-8    &  16  &     128 &  128 GiB \\
        \hline
    \end{tabular}
\end{table}

\pagebreak
Overall, the configuration options in our evaluation dataset do not focus on just scale, since for the most part, that would translate mostly to a cost-performance trade-off~\cite{will2023selecting}.
Instead, the prominent configuration dimensions include the ratio between memory and CPU cores, as well as the distribution of these given resources across fewer large nodes or more numerous but smaller nodes.
This in turn creates a search space of configuration options with different degrees of \emph{efficiency}, i.e., distance from the cost-performance Pareto front.

\vspace{4mm}

\hspace{-4.5mm}
\textit{Job Executions.}\\
Table~\ref{tab:executions} shows the statistical properties of the trace dataset that resulted from executing each of the 18 Spark jobs on each of the ten resource configurations.

\vspace{2mm}
\begin{table}[h]
    \small
    \renewcommand{\arraystretch}{1.4} %
    \centering
    \caption{Statistical Properties of the Evaluation Trace Dataset Containing 180 Spark Job Executions}\label{tab:executions}
    \begin{tabular}{lcc}
        & Cost [USD] & Runtime [seconds]\\
        \hline\\[-3.0ex]
        mean &  1.409 &   1834.832 \\\rowcolor{Gray}
        std  &  2.645 &   2917.467 \\
        min  &  0.177 &   141.680 \\\rowcolor{Gray}
        25\% &  0.457 &   462.730 \\
        50\% &  0.772 &   848.700 \\\rowcolor{Gray}
        75\% &  1.289 &  1722.530 \\
        max  & 26.156 & 21714.740 \\
        \hline
    \end{tabular}
\end{table}
\vspace{2mm}

A job execution cost about \$1.41 USD on average and lasted about half an hour on average.
Note that due to budget constraints, each job was only executed once on each cloud resource configuration, which may make this measured test job data somewhat vulnerable to outliers.
Users of the approach may consider running each test job multiple times to improve the quality of their infrastructure profiling data.

\vfill
\pagebreak
\subsection{Resource Allocation Approaches Tested}\label{resource-allocation-approaches-tested}
\vspace{1.3mm}

As part of our evaluation, we compare Flora to two state-of-the-art baselines and also to several simple baseline approaches.
All approaches attempt to select the most cost-optimal configuration for a given job out of the given configuration options.

Flora is an approach that does not assume the recurrence of a job to learn from runtime data of previous instances of the job or to amortize the initial cost of traversing the configuration search space of the given job.
Consequently, we did not evaluate competing approaches that rely on building performance models from previous instances of a job.
In our evaluation, Flora's computational overhead for each selection is in the millisecond range on consumer hardware.
All other approaches tested also have a low overhead for selecting a resource configuration for an individual given job.

\vspace{2mm}
\hspace{-4.5mm}
The evaluated approaches are as follows:
\vspace{1.0mm}

\begin{itemize}[itemsep=6pt]
    \item \emph{Flora}\\
        This is our method as described in Section~\ref{sec:approach}.
    \item \emph{Flora with one class (Fw1C)}\\
        This approach is similar to Flora, but it skips the classification step and thus learns from every test job and not just from a selection of similar test jobs.
    \item \emph{Juggler}~\cite{al2022juggler}\\
        As highlighted in Section~\ref{sec:related_work}, Juggler is a state-of-the-art approach. It choses configurations for Spark jobs based on providing just enough total cluster memory for in-memory caching after a brief job profiling period. However, Juggler is only applicable for iterative machine learning workloads and is hence only evaluated for K-Means, Linear Regression, and Logistic Regression.
    \item \emph{Crispy}~\cite{will2022get}\\
        As highlighted in Section~\ref{sec:related_work}, Crispy is a state-of-the-art approach. It attempts to configure the cluster according to an estimate of the memory consumption of the given job, which Crispy obtains by extrapolating the memory consumption that it measured during a brief job profiling period.
    \item \emph{Minimizing/maximizing CPU}\\
        This approach means always selecting the configuration option that has the highest/lowest allocation of CPU cores.
    \item \emph{Minimizing/maximizing memory}\\
        This approach is analogous and means always selecting the configuration option that has the highest/lowest allocation of memory.
    \item \emph{Random selection}\\
        This represents the average expected result of a uniformly sampled random choice from all cloud configuration options.
\end{itemize}

\vfill
\pagebreak

\subsection{Cost-Optimized Cloud Resource Selection}\label{ssec:main_results}
\vspace{1.3mm}

In this Section, we present the main results of our evaluation of Flora and the aforementioned baseline approaches.
All approaches attempt to select the most cost-optimal configuration out of the 10 options for each of the 18 jobs in our trace dataset.
We simulate this selection and we then consult our aforementioned evaluation trace dataset to assess to what extent an approach succeeds in identifying the cost optimal configuration option.

For this evaluation we apply GCP's VM pricing as of \mbox{2024-12-01} in Google's Frankfurt data center as the cost model to our dataset.
Further, we normalized the dollar cost and runtime in seconds for each job, so that 1.0 represents the minimum observed value for that given job.
This is to ensure that the quality of each of the 18 configuration selections is weighted equally in this evaluation.

\vspace{2mm}
\begin{table}[htb]
    \small
    \renewcommand{\arraystretch}{1.4} %
    \centering
    \caption{Results of Cloud Resource Selection Approaches,\\Normalized to 1 $=$ Optimal, then Averaged Over All 18 Jobs.}\label{tab:main_results}
    \begin{tabular}{lcc}
        \text{Approach} & \text{Cost} & \text{Runtime} \\
        \hline\\[-3.0ex]
           minimize CPU       &        2.126     &      7.837  \\\rowcolor{Gray}
       random selection       &        1.941     &      3.484  \\
        minimize memory       &        1.864     &      3.166  \\\rowcolor{Gray}
           maximize CPU       &        1.590     &  \gn{1.346} \\
        maximize memory       &        1.487     &      1.442  \\\rowcolor{Gray}
        Flora with one class  &        1.336     &      1.952  \\
                Juggler       &        1.334     &      2.973  \\\rowcolor{Gray}
                  Flora       &    \gn{1.052}    &      1.578  \\
                  \hline
    \end{tabular}
\end{table}
\vspace{4mm}

Table~\ref{tab:main_results} presents a summary of this evaluation.
Here, we averaged the normalized cost and runtime values over all 18 jobs.
We see that Flora's selections achieve a mean normalized cost of 1.052, meaning that, on average, the cloud configuration that Flora selects for a given job merely incurs a 5.2\% higher execution cost than what the cheapest available choice would have incurred.
This is by far the lowest value out of all tested approaches.

The mean normalized runtime Flora achieves with these selections is 1.578, which means that, on average, Flora chooses a configuration that leads to an execution duration 57.8\% longer than what would have been possible with the fastest configuration option for that given job.
Comparatively, this means that Flora's cost-optimized selections still lead to decent performance, only being beaten by the baseline approach that always chooses the configuration with the highest total CPU allocation.
This suggests that the low cost is achieved in part by occupying the efficiently used resources for shorter durations.

\vfill
\pagebreak

\newcommand{\hs}{\hspace{1.5mm}}

\begin{table}[htb]
    \centering
    \renewcommand{\arraystretch}{1.2} %
    \scriptsize
    \caption{Configuration Selection and Resulting Normalized Job Execution Cost for Each Job, with 1 = Optimal}\label{tab:details}
    \begin{tabular}
        {
            l@{\hskip 1.9mm}r@{\hskip 1.3mm} | r@{\hskip 2mm}r@{\hskip 1.3mm} |r@{\hskip 2mm}r@{\hskip 1.3mm} | r@{\hskip 2mm}r@{\hskip 1.3mm} | r@{\hskip 2mm}r@{\hskip 1.3mm}
        }
        \multicolumn{2}{c}{} & \multicolumn{2}{c}{\footnotesize Crispy} & \multicolumn{2}{c}{\footnotesize Juggler} & \multicolumn{2}{c}{\footnotesize Fw1C} & \multicolumn{2}{c}{\footnotesize Flora}  \\
        \hline\\[-2.5ex]
        {\footnotesize Grep           }    & 3010 GiB & \#7 & 1.711      &  -- &   {--  \hs}  & \#9 &     1.381  & \#1   & \gn{1.000}  \\
        {\footnotesize Grep           }    & 6020 GiB & \#7 & 1.730      &  -- &   {--  \hs}  & \#9 &     1.421  & \#1   & \gn{1.000}  \\\rowcolor{Gray}
        {\footnotesize GroupByCount   }    &  280 GiB & \#2 & 1.389      &  -- &   {--  \hs}  & \#9 &     1.445  & \#1   & \gn{1.000}  \\\rowcolor{Gray}
        {\footnotesize GroupByCount   }    &  560 GiB & \#3 & 1.870      &  -- &   {--  \hs}  & \#9 &     1.423  & \#1   & \gn{1.003}  \\
        {\footnotesize Join           }    &   85 GiB & \#9 & \gn{1.196} &  -- &   {--  \hs}  & \#9 & \gn{1.196} & \#9   & \gn{1.196}  \\
        {\footnotesize Join           }    &  172 GiB & \#9 & \gn{1.093} &  -- &   {--  \hs}  & \#9 & \gn{1.093} & \#9   & \gn{1.093}  \\\rowcolor{Gray}
        {\footnotesize K-Means        }    &  102 GiB & \#7 & 1.482      & \#7 &     1.482    & \#8 &     1.308  & \#9   & \gn{1.237}  \\\rowcolor{Gray}
        {\footnotesize K-Means        }    &  204 GiB & \#2 & \gn{1.000} & \#2 & \gn{1.000}   & \#8 &     2.158  & \#9   &     1.081   \\
        {\footnotesize Lin. Regression}    &  229 GiB & \#2 & \gn{1.000} & \#7 &     1.503    & \#9 &     1.053  & \#9   &     1.053   \\
        {\footnotesize Lin. Regression}    &  459 GiB & \#3 & \gn{1.076} & \#2 &     1.294    & \#9 &     1.146  & \#9   &     1.146   \\\rowcolor{Gray}
        {\footnotesize Log. Regression}    &  210 GiB & \#3 & 1.066      & \#2 &     1.435    & \#9 & \gn{1.045} & \#9   & \gn{1.045}  \\\rowcolor{Gray}
        {\footnotesize Log. Regression}    &  420 GiB & \#3 & 1.292      & \#3 &     1.292    & \#9 & \gn{1.000} & \#9   & \gn{1.000}  \\
        {\footnotesize SelectWhereO...}    &   92 GiB & \#3 & 1.772      &  -- &    {-- \hs}  & \#9 &     1.334  & \#1   & \gn{1.000}  \\
        {\footnotesize SelectWhereO...}    &  185 GiB & \#7 & 1.496      &  -- &    {-- \hs}  & \#9 &     1.307  & \#1   & \gn{1.000}  \\\rowcolor{Gray}
        {\footnotesize Sort           }    &   94 GiB & \#2 & 1.251      &  -- &    {-- \hs}  & \#2 &     1.251  & \#9   & \gn{1.050}  \\\rowcolor{Gray}
        {\footnotesize Sort           }    &  188 GiB & \#2 & 1.941      &  -- &    {-- \hs}  & \#2 &     1.941  & \#9   & \gn{1.031}  \\
        {\footnotesize Word Count     }    &   39 GiB & \#9 & 1.258      &  -- &    {-- \hs}  & \#9 &     1.258  & \#1   & \gn{1.000}  \\
        {\footnotesize Word Count     }    &   77 GiB & \#9 & 1.294      &  -- &    {-- \hs}  & \#9 &     1.294  & \#1   & \gn{1.000}  \\
        \hline\hline\\[-2.5ex]
        {\footnotesize Mean           }   &          &     & 1.384      &     & 1.334        &     & 1.336      &       & \gn{1.052}  \\
    \end{tabular}
\end{table}

Table~\ref{tab:details} details the selection quality of Flora and the main baseline approaches for each of the 18 jobs.
We see that Flora's cloud resource selections generally lead to optimal or near-optimal execution cost, overall outperforming baseline approaches.

Out of the ten available cloud configuration options, Flora ended up choosing configuration \#9 for all jobs of class~\texttt{A}.
This indicates that any combination of four out of the five test jobs of class \texttt{A}, on average, have seen the most cost-optimal solution with configuration \#9.
While configuration \#9's total cluster memory of 256 GiB allows for caching at least parts of the input datasets of the memory-demanding jobs of class \texttt{A}.
Meanwhile, it is notable that Flora made a clear choice in favor of spreading the total cluster CPU cores and memory over 16 smaller nodes as compared to configuration \#2, which has the same number of CPU cores and memory, but spread over just 8 larger nodes.
This suggests that the jobs of this class do not just share a preference for the allocation of memory for caching, but they also share a preference for the distribution of the resources among cluster nodes.

In line with the assertion that jobs of class \texttt{B} do not need significant amounts of memory for caching large parts of the dataset, Flora chose \#1 configuration for those jobs.
This is the configuration that has the least total cluster memory and resulted in cost-optimal execution for almost all class~\texttt{B} jobs.

\subsection{Influence of the Resource Cost Structure}

The total cost of an execution is determined by the execution duration and the hourly cost of the resources used during the execution.

Once the cost of individual cloud resources changes, the cost-efficiency ranking of the available cloud configurations may change as well.
For instance, fluctuating renewable energy production continuously changes the environmental cost of operating certain cloud resources~\cite{wiesner2021let}.
Further, fluctuating demand and availability due to shared infrastructure use vary the cost of using individual cloud resources, either in terms of price in public clouds~\cite{lee2022spotlake} or capacity consumption in private clouds.

\begin{figure}
  \centering
    \includegraphics[width=\columnwidth, keepaspectratio]{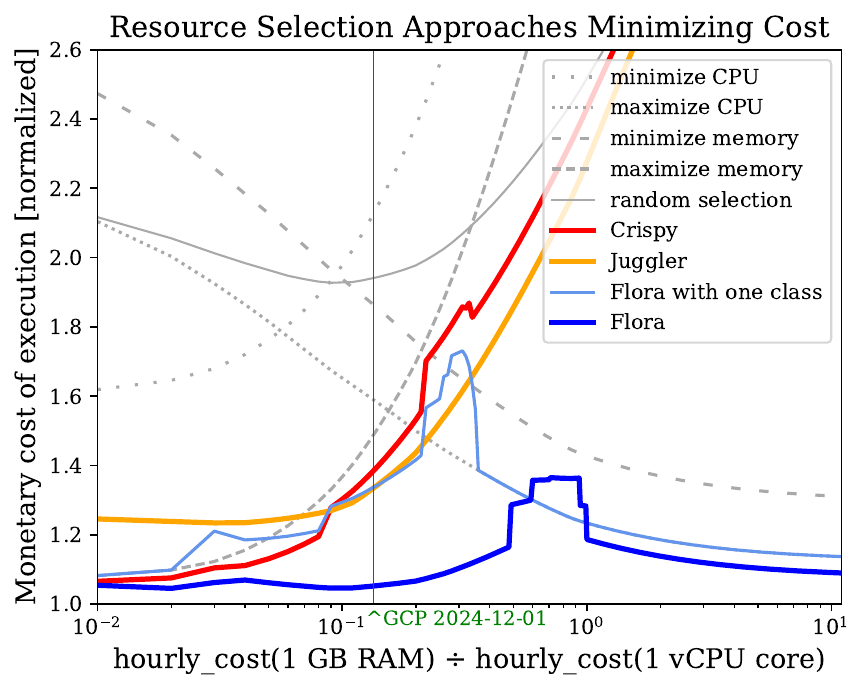}
  \caption{Comparing cloud configuration selection approaches for varying individual resource costs.
Main approaches are highlighted in color, while basic baselines for reference are shown in gray.}\label{fig:prices}
    \vspace{-4mm}
\end{figure}

In Figure~\ref{fig:prices}, we investigate how the quality of resource selections of Flora and baseline approaches change in response to changes in the cost structure of the available resource options.
In particular, we focus here on the prices of 1 GB of memory and 1 vCPU core.
On the far left of the x-axis (\(10^{- 2}\)), the hourly cost of 1 GB of memory is equal to the hourly cost of 0.01 vCPU cores.
On the far right of the x-axis (\(10^{1}\)), the hourly cost of 1 GB of memory is equal to the hourly cost of 10 vCPU cores.
For reference, we mark the price points we used in the main evaluation experiment in Section~\ref{ssec:main_results} with a thin green vertical line.

We see that Flora is able to react to price changes in individual resources comparatively well.
The pricing points where Flora adjusts its resource selection correspond to the points where the plot line contains a ``step''.
Meanwhile, approaches that directly allocate individual resources only work as long as the hourly cost of the resource is justified by the performance gained from that resource.
This can be observed with Juggler and its memory allocation for in-memory caching, as well as the approaches minimizing or maximizing the allocation of a particular resource.

In our experiments, the hourly costs for resource configurations with the same amount of total cluster memory and total cluster CPU cores are equal, regardless of scale-out.
While this represents the actual cost model for the cloud resources we used to generate the trace dataset as of 2024-12-01, it may differ for other infrastructures or cost denominations other than price.
However, conceptually, this cannot break Flora either, since its selection decision is not based on the current cost of individual aspects of the resource configuration, but based on the current cost of discrete configuration options.

\subsection{Influence of Flora's Classification Accuracy}\label{ssec:accuracy}

Flora uses a classification step to select which test jobs to learn from, preferring ones that are similar to the given job.
In this section, we investigate how the classification accuracy for jobs affects the quality of resulting resource selection.

We repeated the main evaluation of Section~\ref{ssec:main_results}, but varied the number of given jobs that we misclassify on purpose to simulate user annotation by non-experts.
The test jobs' classification remains accurate in this experiment, assuming it has been done by experts, since this classification only needs to be done once per test job for all users of the approach.

\begin{figure}[h]
  \centering
  \includegraphics[width=\columnwidth, keepaspectratio]{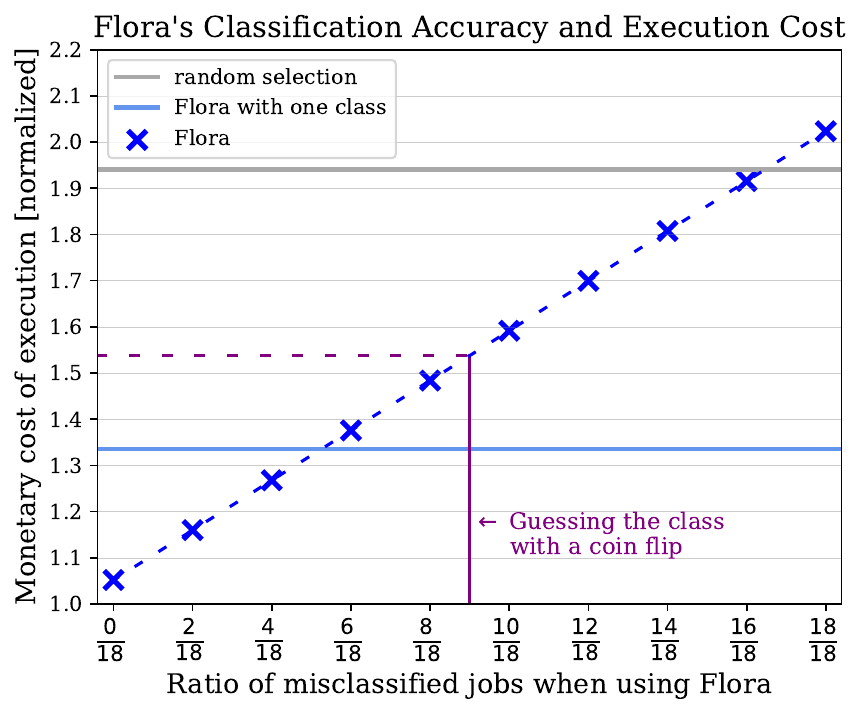}
  \caption{Evaluating Flora's resource selection quality for varying job classification accuracy.}\label{fig:accuracy}
\end{figure}

Figure~\ref{fig:accuracy} shows the results of this experiment.
We see that users who misclassify a third or more of their given jobs would achieve better results using Flora with a single class instead of the original method with two classes.
Furthermore, we can observe that Flora users who guess their job's class with a coin flip are still expected to achieve significantly better results than users that randomly select a cloud configuration themselves, without using Flora.

\subsection{Discussion}\label{ssec:discussion}

The evaluation results have demonstrated that jobs with similar data access patterns share similar preferences for cluster resources, especially with respect to memory allocation.
Meanwhile, suitable memory allocation has a significant impact on the cost efficiency of a job execution, as been found in related work~\cite{al2022juggler,al2022blink,will2022get,will2023selecting}.

For the investigated jobs and the cloud configurations, Flora's binary decision in favor of allocating memory appears to yield better results than directly estimating the right amount of memory to allocate, as attempted by, e.g., Juggler~\cite{al2022juggler}.
On the one hand, this may be due to the coarse-grained total cluster memory options within the available cloud configuration options of our trace dataset.
On the other hand, it may be due to the fact that Flora automatically favors cost-efficient cloud configurations in general, e.g., by automatically choosing an appropriate distribution of the allocated resources across the cluster nodes.

Our approach has negligible computational overhead beyond a one-time infrastructure profiling phase that is unique per infrastructure offering, and the results can be utilized by all users of the distributed dataflow system on said infrastructure.
For instance, the method could be implemented by cloud providers, who conduct the infrastructure profiling to offer all of their users cost-efficient processing.

Without needing to repeat the infrastructure profiling step, users could re-implement the job classification aspect, e.g., further subdivide or merge classes.
As seen when comparing Flora with one vs.
with two classes, a more finegrained classification may lead to more accurate job behavior predictions, but it also increases the possibility of misclassifications.

Non-expert users, without sufficient knowledge of their job, can fall back to a one-class version of Flora and still achieve reasonable cost efficiency for the first execution of their job.
Meanwhile, if their job happens to be recurring, the more suitable class in a two-class approach could be found automatically by trying both class assumptions for the same job and examining which class assumption leads to a lower cost.
This search cost would then be amortized by subsequent instances of the recurring job.

\section{Related Work}\label{sec:related_work}

In this section, we discuss related approaches to resource allocation for distributed dataflows.
We group the works by what information the approaches base their decision on.
Related approaches to our research problem learn job behavior from historical job executions, from dedicated job profiling, infrastructure profiling, or a combination of multiple information sources.

\subsection{Learning from Historical Executions}

Several approaches use historical runtime data to predict a job's performance.
This runtime data is typically obtained from full previous executions of a recurring job~\cite{rajan2016perforator,hongzi2016resource,chen2021silhouette,cherrypick,hsu2018arrow,hsu2018micky}.
The models are then used to predict the execution performance for different cluster configurations, allowing to choose an optimized configuration.

\emph{CherryPick}~\cite{cherrypick} and related approaches~\cite{hsu2018micky,hsu2018arrow}, for example, use Bayesian optimization to iteratively predict the cost-optimal cloud configuration based on all available historical executions, while respecting a minimum performance target.
The exploration phase ends when the cost of further search is deemed unjustified.

Meanwhile, \emph{Silhouette}~\cite{chen2021silhouette} is a cloud configuration selection method based on performance modeling with low training overhead.
Silhouette uses a model transformer for rapid transfer learning, and can optimize cloud configurations under constraints.

The drawback of all resource selection approaches based on historical executions is that they only work for jobs that are recurring in some form, without solving the difficult problem of distinguishing historical jobs that behave too different to be considered the same job still.
At the same time, Microsoft and Alibaba, for instance, have identified only 40-65\% of jobs in their data centers as recurring~\cite{agarwahl2012reoptimizing,jyothi2016morpheus,wang2020grosbeak}.

Unlike these approaches, Flora does not assume the existence of any previous or subsequent executions of the given job.
This also allows Flora to avoid the problem of determining when a job is deemed similar enough to be an instance of a recurring job, considering possible changes in input dataset characteristics or job parameters which may influence the job's resource access patterns significantly.

\subsection{Learning from Job Profiling}

Several other approaches attempt to efficiently learn enough information about just the given job to make a resource allocation decision~\cite{ernest,al2022juggler,al2022blink,will2022get,baughman2018profiling,popescu2013predict}.

\emph{Ernest}~\cite{ernest} trains a parametric model for the scale-out behavior of jobs from the results of profiling runs on reduced input data, which works well for data-parallel programs that exhibit intuitive scale-out behavior.
Ernest performs profiling runs on the target infrastructure, leading to relatively high overhead in cost and time.

\emph{Juggler}~\cite{al2022juggler} and \emph{Blink}~\cite{al2022blink} by Al-Sayeh et al. can use small job profiling runs independent of the target infrastructure to measure the ratio between input dataset size and cached data size for Spark jobs.
This information is then used to allocate cluster resources with sufficient memory to enable in-memory caching of the dataset.

\emph{Crispy}~\cite{will2022get}, our own work, follows a similar approach as Juggler and Blink by estimating the memory required for a full job execution, but differs in details of how memory usage is measured during profiling.
Like Juggler and Blink, Crispy has only been shown to work well on some jobs, notably ones that exhibit a straightforward relation between input sizes and memory use, e.g., for caching.

These job-profiling based approaches, like Flora, do not assume job recurrence.
Thus, they try to configure the cluster directly, without relying on performance data from full historical executions, i.e., previous instances of a recurring job.
In contrast, however, Flora can utilize knowledge gained from executions of loosely related historical jobs on the target infrastructure.

\subsection{Learning from Infrastructure Profiling}

Some related approaches use infrastructure profiling to learn about available configuration options and apply this knowledge during their configuration selection process.

\emph{RUPAM}~\cite{xu2018heterogeneity}, is a heterogeneity-aware task scheduling system for big data platforms, which considers both task-level resource access patterns characteristics as well as available hardware characteristics.
It expands Spark's native scheduling decisions which already respect data locality concerns that are rather characteristic for co-located compute and storage nodes of on-premise clusters.
It uses \emph{sysbench} and \emph{iperf} to measure CPU, I/O, and network performance of the available machines.
While RUPAM aims to maximize performance, given the available resource options, it does not consider the operating cost of different resources and is therefore not suitable for cloud-based distributed dataflow job execution.

\emph{Perona}~\cite{scheinert2022perona}, our own work, employs an infrastructure fingerprinting approach that automatically selects statistically significant benchmarking metrics and saves the gathered benchmarking data in a public repository.
Users of the approach can then utilize the infrastructure fingerprints to iteratively find an optimized configuration quicker than standalone iterative search based configuration approaches like Arrow or CherryPick.
Perona therefore also shares these approaches' limitation of needing to expend significant computational power to choose a resource configuration for each new job, thus also relying on the assumption that the main runtime-behavior-influencing factors remain consistant between instances of a recurring job.

Unlike these approaches, Flora uses actual distributed dataflow jobs as test jobs to profile the infrastructure.
Also, Flora has no need for an adaptation step for each new given job and does not rely on any test runs of the given job itself to make a configuration selection.
Instead, the given job undergoes a user-annotated classification.

\section{Conclusion}\label{sec:conclusion}

In this paper, we presented Flora, a new approach to selecting cloud resources for cost-optimized big data processing.
We have shown that Flora selects near-optimal resource configurations and, after an initial infrastructure profiling step, imposes virtually no resource selection overhead.
Flora accomplishes this by leveraging categories of jobs, treating a given job as either memory-demanding or memory-yielding, and then selecting the cloud configuration that was optimal for a category of jobs in an infrastructure-specific profiling trace, respecting current cloud resource costs.

In the future, we want to automate and improve the classification step of our approach to help users submit jobs they have limited knowledge about.
In particular, we plan to explore a combination of static code analysis and minimal profiling of given jobs.

\section*{Acknowledgments}

This work has been supported through grants by the German Research Foundation (DFG) as ``C5'' (Project 506529034) and ``FONDA'' (Project 414984028, SFB 1404), and by the Google Cloud Research Credits Program.

\bibliographystyle{IEEEtran}
\balance
\bibliography{./references}

\end{document}